\documentclass[a4paper,11pt]{article}
\usepackage{jinstpub}
\usepackage{lineno}

\title{\boldmath FPGA-based muon shower identification and graph neural network tracking algorithms for HL-LHC triggers}

\author[1]{D. Estrada,\note{Corresponding author.}}
\author{E. Aller,}
\author{A. Cardini,}
\author{S. Folgueras,}
\author{P. Leguina,}
\author{and J. Prado}
\affiliation{University of Oviedo,\\C. San Francisco 3, Oviedo, Spain}
\affiliation{Instituto de Ciencias y Tecnologías Espaciales de Asturias (ICTEA),\\ C. de la Independencia 13, Oviedo, Spain}

\emailAdd{destrada@cern.ch}

\abstract{This work presents two hardware-accelerated strategies for the upgrade of the CMS Level-1 Trigger (L1T) to target efficiency losses due to highly energetic radiating muons and unconventional physics signatures, such as Long-Lived Particles (LLPs). Operating directly on low-level detector hits, the first strategy implements a dedicated muon shower identification algorithm in the CMS barrel region that monitors hit multiplicities in the Drift Tube chambers to generate Shower Primitives. Executing with sub-bunch-crossing latency and ultra-low resource consumption, this algorithm successfully tags showered muons and provides essential context for downstream track finders. The second approach targets the barrel-endcap transition (overlap) region by evaluating Graph Neural Networks to reconstruct displaced muon tracks. Following a divided methodology that first isolates hardware feasibility from evolving physics refinements, an initial proxy model based on a GraphSAGE architecture is evaluated to validate its viability on FPGAs within strict latency limits. This implementation utilizes an INT8-PO2 quantization technique coupled with a data-driven bit-width optimization. By replacing DSP-heavy fixed-point multiplications with fast compile-time arithmetic bit-shifts and allocating the minimum required bits per signal, this hardware-software co-design reduces DSP utilization to $20\%$ and achieves a deterministic inference latency of just $19$ clock cycles. Together, these developments pave the way for triggering on non-standard muon signatures, such as those from LLP decays, at L1T.}  

\begin{document}
\maketitle
\flushbottom

\section{Introduction}
\label{sec:intro}
To meet the data demands of the upcoming High-Luminosity Large Hadron Collider (HL-LHC), the Compact Muon Solenoid (CMS) experiment is deploying enhanced Level-1 trigger (L1T) architectures built on next-generation hardware. This infrastructure is engineered to handle several tens of Tb/s with ultra-low latency of $\mathcal{O}(\mu \text{s})$, allowing complex, offline-style algorithms that were previously beyond the computational reach of the online trigger stage to be deployed directly onto Field-Programmable Gate Arrays (FPGAs). Leveraging this processing power makes it feasible to target rare phenomena early at the trigger level. For instance, Long-Lived Particles (LLPs), predicted by many Beyond the Standard Model (BSM) scenarios, produce displaced vertices, delayed signals, and unusual energy deposits that conventional triggers---which are traditionally optimized for prompt signatures---often miss~\cite{CMS:2020, Zabi:2022}.

We aim to bridge this gap by developing intelligent trigger strategies that embed machine-learning algorithms and advanced pattern recognition into the ultra-fast hardware of the CMS Level-1 muon trigger, enabling sensitivity to these elusive signatures at the earliest stage of data acquisition \cite{Folgueras:2025}.

\begin{figure}[htbp]
\centering
\includegraphics[width=\textwidth]{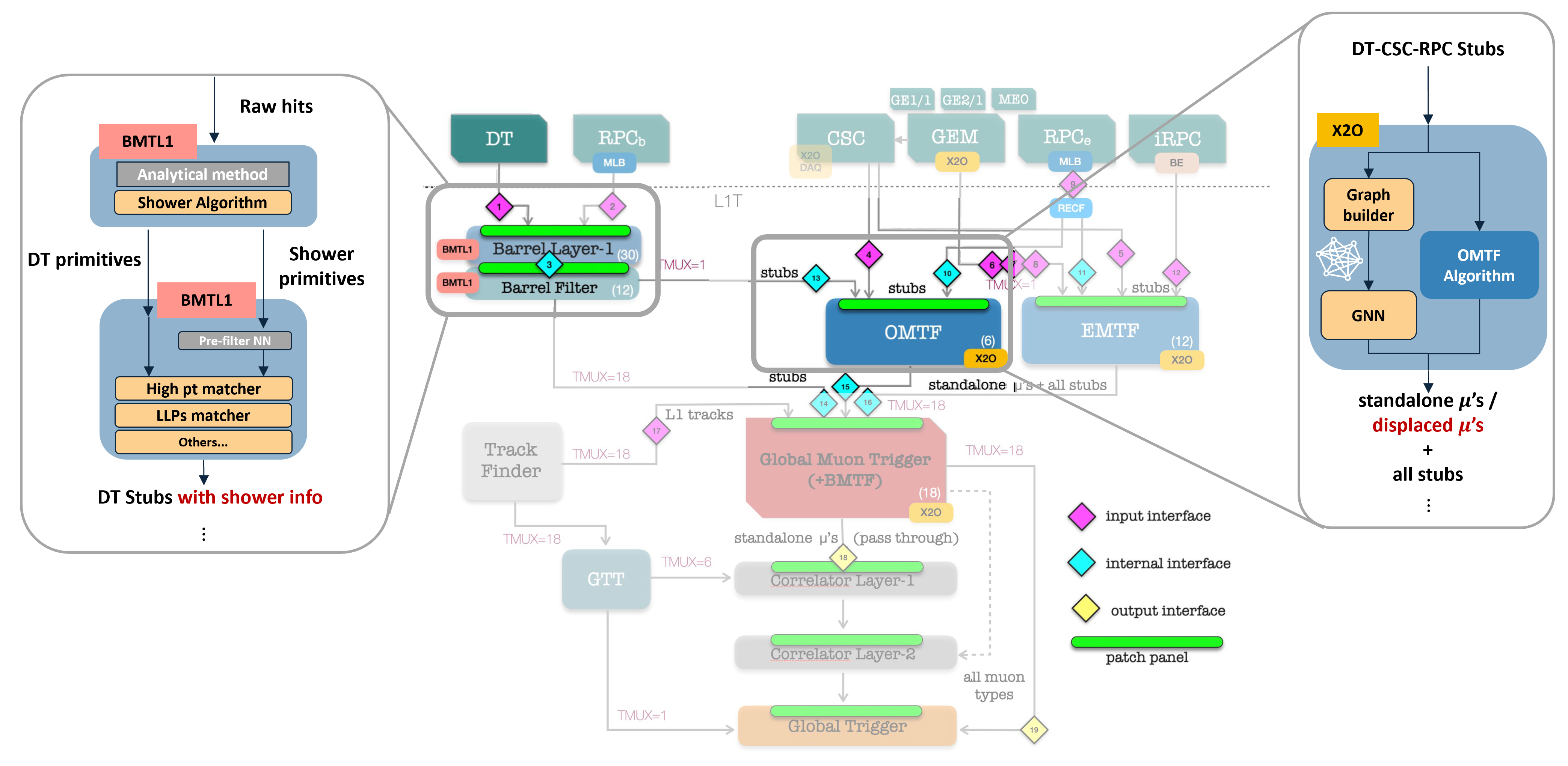}
\caption{Schematic of the CMS phase2-L1T architecture highlighting the two specific contribution areas related with this work. The left inset details a muon shower identification algorithm in the Barrel region. The right inset shows a GNN approach integrated in parallel with the OMTF.\label{fig:cms-trigger}}
\end{figure}

This research focuses on two complementary strategies operating directly on low-level hit information from the muon system (see figure \ref{fig:cms-trigger}). The first approach is the development of a muon shower identification algorithm in the CMS Barrel region. By monitoring hit multiplicities in the Drift Tube (DT) chambers, this algorithm aims to recover trigger efficiency losses caused by high-momentum muons---which often undergo radiative energy losses that confuse standard tracking algorithms \cite{Sirunyan:2020}---and to potentially identify LLP signatures within the barrel. Section~\ref{sec:muon-showers} describes the implementation and performance of this muon shower identification algorithm. The second strategy employs Graph Neural Networks (GNNs) in the overlap region between the barrel and endcap detectors of CMS to reconstruct displaced muon tracks. Regarding the latter, Section~\ref{sec:gnn} discusses the development of this GNN-based track finding approach for the overlap muon track finder (OMTF), specifically presenting the advances in validating the feasibility of executing spatial message-passing GNNs within the strict latency windows of the hardware trigger using an initial GraphSAGE proxy model. Finally, Section~\ref{sec:conclusions} provides concluding remarks and outlines future directions for both strategies.

\section{Muon shower identification algorithm}
\label{sec:muon-showers}
The reconstruction of muons in the CMS L1T relies heavily on cross-matching trigger primitives (TPs)/stubs generated by each muon sub-detector: DT chambers, Cathode Strip Chambers (CSCs), Gas Electron Multipliers (GEMs), and Resistive Plate Chambers (RPCs) \cite{CMS:2020, Zabi:2022}. In the barrel region specifically, these stubs are generated by the Analytical Method (AM) algorithm \cite{Abbiendi:2023}, which fits muon paths as straight lines based on combinations of DT hits. However, highly energetic muons often undergo radiative energy losses, such as bremsstrahlung, when traversing the steel of the magnet flux-return yoke, creating localized electromagnetic showers. These showers generate a combinatorial explosion of hits in the DT chambers, causing ambiguities for standard pattern recognition algorithms and leading to a degradation of the L1T efficiency of up to $10\%$ for high-momentum muons \cite{Sirunyan:2020}. Furthermore, similar anomalous hit clusters can be produced by the decays of LLPs beyond the inner region of CMS \cite{Cottin:2023, Syrunyan:2021}.

To mitigate these efficiency losses and target unconventional signatures, a dedicated shower identification algorithm has been developed to operate directly on DT raw hits. This algorithm is planned to run in parallel with the AM algorithm at the local trigger stage on the BMTL1 hardware board \cite{CMS:2020}. The core logic functions by monitoring hit multiplicities within a $400\text{ ns}$ time window (equivalent to 16 LHC bunch crossings---BXs) to account for the maximum electron drift time within the DT chambers. When the accumulated number of hits exceeds a predefined, adjustable threshold, a compact Shower Primitive (SP) is generated. This SP encodes spatial and timing information such as the BX of the first detected hit, the minimum and maximum hit identifiers, and a vector representing the frequency profile of the activated hits (i.e., the shower’s hit density) \cite{Acevedo:2026, Prado:2026}. Rather than replacing standard tracking, this information is intended to support it.

In the subsequent trigger stage, called Barrel Filter, a second set of BMTL1 boards will be used to concentrate the information across all $5$ wheels and up to $3$ of the $12$ $\phi$-sectors of the muon barrel system (with 12 overlapping Barrel Filter boards covering the full cylinder and providing full redundancy) \cite{CMS:2020}. At this stage, other correlation algorithms can be applied before moving the AM TPs to the barrel muon track finders (BMTF). In this framework, an algorithm to identify hihg-energy muons is under development, it consists of a geometrical matching to evaluate whether standard AM stubs intersect with the spatial segments defined by the SPs. The current version performs the matching by formulating the stubs and shower segments as cartesian linear equations to determine their exact intersection using Cramer's rule \cite{Acevedo:2026, Prado:2026}. This algebraic approach is specifically chosen for hardware deployment because it reduces the intersection problem to simple determinant evaluations---relying purely on basic multiplications and subtractions---that are highly efficient to execute within the strict resource constraints of an FPGA. By tagging AM primitives that point toward showered regions, downstream track finders would be provided with context to handle potential track losses and prevent the misreconstruction of kinematics.

Evaluations based on simulated data demonstrate that the described approach is effective. The shower primitive production algorithm achieved an accuracy of around $80\%$ while maintaining a local trigger rate under HL-LHC conditions that is approximately ten times lower than that of the standard AM algorithm, and the high-energy muon tagging implementation successfully achieved efficiencies greater than $80\%$. 

The hardware implementation fits within the L1T constraints. The initial base shower generation features ultra-low resource consumption, utilizing only around $2.5\text{k}$ Lookup Tables (LUTs) and $2.5\text{k}$ Flip-Flops (FFs), and executes with a sub-BX latency of approximately $20\text{ ns}$. To complement this, the high-energy matching block requires minimal overhead. While a full-scale firmware version is under study, a preliminary draft capable of correlating information across $3$ sectors and $1$ wheel was achieved using only $4.3\text{k}$ LUTs and $5.6\text{k}$ FFs. This matching block can process $48$ TP-SP pairs per clock cycle at $480 \text{MHz}$, effectively operating within the sub-BX latency margin. These figures show that the algorithm can run in real time without affecting the fixed-latency trigger pipeline.

The base shower generation design was successfully deployed and integrated into CMS data acquisition during recent slice tests to evaluate its performance with real cosmic data before the long shutdown period started.

\section{Graph Neural Networks for the Overlap Muon Track Finder}
\label{sec:gnn}
Within the CMS L1T, the barrel-endcap transition region ($0.83<|\eta|<1.24$) is particularly challenging as the OMTF must combine stubs from DT, CSC, and RPC sub-detectors in an inhomogeneous magnetic field \cite{CMS:2020, Zabi:2022}. The current OMTF algorithm relies on matching these hits to predefined reference patterns based on a naive Bayes classifier \cite{BUNKOWSKI:2019, Zabolotny:2016}, an approach that becomes suboptimal for non-standard trajectories such as displaced muons originating from LLPs. To overcome this limitation, GNNs are being developed to exploit the full detector granularity \cite{Leguina:2025, Cardini:2026}.

The stubs are represented as graph nodes, with edges connecting stubs that are geometrically compatible in $\eta-\phi$. Each edge carries normalised attributes (layer,$\eta$, $\phi$, $\Delta\eta$, and $\Delta\phi$). Through message passing, the GNN assigns a compatibility score to each stub pair, separating genuine track segments from pile-up combinations. This representation naturally handles the sparse, irregular occupancy of the detector and is well suited to non-standard topologies. Initial studies have suggested that architectures based on graph sample-and-aggregate operations (GraphSAGE) \cite{Hamilton:2017} or more general Message Passing Neural Networks (MPNNs) could provide strong performance for this target use, consistent with prior demonstrations of GNN inference within sub-microsecond FPGA latency budgets for real-time particle reconstruction \cite{Iiyama:2020}.

However, executing this type of spatial message-passing GNNs on FPGAs within the strict nanosecond-scale latency limits of the hardware trigger is challenging because of irregular data dependencies and the cost of multiplications in Digital Signal Processing (DSP) resources. To address this issue, a divided methodology has been adopted: first, evaluating the hardware feasibility of deploying such models within the strict FPGA resource budgets, and subsequently refining the model architectures to solve the actual physical problem. This work focuses exclusively on presenting the advances of this first hardware-validation stage. To isolate synthesis and timing feasibility from the evolving physics simulation, a hardware-tailored methodology was established using a decoupled firmware-proxy benchmark based on a two-layer GraphSAGE architecture, evaluated using subgraphs from the standard Cora dataset (a benchmark graph where nodes represent scientific publications connected by citation edges) \cite{Sen:2008}. This proxy utilizes a fixed-size 2-hop subgraph that mirrors the typical node multiplicity found in CMS muon tracks \cite{Cardini:2026}. 

Typically, feature updates in message-passing networks explicitly include the root-node self-term alongside aggregated neighbor features to compute the new representations \cite{Hamilton:2017}. However, in this hardware-oriented approach, we deliberately opted to omit this self-connection. By making this structural choice, an entire weight matrix is removed from the datapath, which eliminates the need for intermediate concatenation buffers and simplifies the fixed-size loop structure required for complete array unrolling
during High-Level Synthesis (HLS) \cite{Cardini:2026}.

To further align the model with the rigorous resource budgets of the trigger hardware, a comprehensive end-to-end optimization workflow was established. Starting from a floating-point PyTorch model, the workflow comprises fixed-size graph extraction, integer-only post-training quantization, power-of-two (PO2) scale approximation, data-driven bit-width optimization, and bit-exact validation against a Python integer reference \cite{Cardini:2026}. The two most critical components of this methodology are detailed below.

In standard integer quantization (INT8), the accumulated results of a neuron's calculation must be rescaled to prevent data overflow before passing to the next layer \cite{Jacob:2018}. Mathematically, this standard rescaling is expressed as:
\begin{equation*}
    q_{\text{out}}  = \text{clamp}(acc\cdot\alpha \gg M),
\end{equation*}

where $acc$ is the $32$-bit accumulator, $\alpha$ is a non-integer scaling factor derived during quantization, and $M$ is a bit-shift used to adjust the binary point. On an FPGA, multiplying by a non-integer factor ($\alpha$) explicitly requires dedicated DSP blocks, which are a scarce resource and introduce significant processing delays. This work proposes to overcome this bottleneck by introducing an INT8-PO2 optimization. Instead of using exact scaling factors, this technique approximates them to the nearest power of two, such that $\alpha\approx 2^p$ \cite{Cardini:2026}. By adopting this approximation, the operation is mathematically reduced to a single bit-shift:

\begin{equation*}
    q_{\text{out}} = \text{clamp}( acc \gg k ),
\end{equation*}

where the shift constant is $k = M - p$. This rescaling step requires no DSPs. The residual DSP usage comes from the matrix multiplications.

The workflow also applies a data-driven bit-width optimization. Rather than using uniform or worst-case precision across the entire network, this technique analyzes the actual numerical distribution of the data to assign the minimum required bits for each signal, preventing overflow while avoiding unnecessary resource allocation. This strategy reduces the overall memory footprint and arithmetic complexity of the hardware design, generating optimized HLS constants per layer \cite{Cardini:2026}.

With this approach, DSP utilization drops from $56\%$ in standard fixed-point post-training quantization (Fixed-PTQ) and $55\%$ in standard quantization-aware training (QAT) down to $20\%$ DSP slices, along with $6\%$ FFs, and $27\%$ LUTs, while maintaining a baseline accuracy of $(75.0 \pm 1.1)\%$. Coupled with data-driven bit-width narrowing,  this multiplier-free rescaling achieves an ultra-low deterministic inference latency of $19$ clock cycles (approximately $52.8 \text{ns}$ at the nominal $360 \text{MHz}$ clock) compared to $77$ cycles for Fixed-PTQ and $55$ cycles for QAT, while demonstrating bit-exact consistency between Python software emulation and RTL simulation \cite{Cardini:2026}.

\section{Conclusions and perspectives}
\label{sec:conclusions}
This work demonstrates that embedding machine learning and advanced pattern recognition directly into the FPGAs of the CMS L1T is a highly effective strategy to meet stringent latency constraints while enhancing the discovery potential of the experiment.  

The muon shower identification algorithm designed for the CMS Barrel region provides a hardware-friendly solution to mitigate trigger efficiency losses caused by highly energetic radiating muons. Consuming minimal FPGA resources and executing with sub-BX latency, the base algorithm was successfully integrated into CMS data acquisition during recent slice tests. Moving forward, the high-energy matching block is being optimized by integrating longitudinal ($\theta$) primitives. Furthermore, future studies will evaluate how shower primitives can be used to identify displaced vertices, such as looking for two close-by showers in the muon stations, which could serve as a direct signature of an LLP decay into jets. 

In parallel, the development of GNNs for the OMTF has shown that spatial message-passing architectures can be executed within the strict latency and resource boundaries of the L1T. The evaluation of the initial GraphSAGE proxy model established that quantization using the INT8-PO2 optimization reduces DSP usage and latency substantially while maintaining an accuracy of $(75.0 \pm 1.1)$\%. Building upon this foundational firmware architecture, ongoing research is shifting toward physics-driven datasets for displaced-muon reconstruction and the deployment of more complex architectures aiming to perform direct edge classification, charge identification, and precise transverse momentum regressions for displaced muons directly in the hardware.

\section*{Acknowledgements}

This work is supported by ERC grant INTREPID, 10.3030/101115353. Funded by the European Union. Views and opinions expressed are however those of the author(s) only and do not necessarily reflect those of the European Union or the European Research Council Executive Agency. Neither the European Union nor the granting authority can be held responsible for them.

\bibliographystyle{JHEP}
\bibliography{biblio.bib}

@techreport{CMS:2020,
    author      = "{CMS Collaboration}",
    title       = "{The Phase-2 Upgrade of the CMS L1 Trigger Technical Design Report}",
    institution = "CERN",
    number      = "CERN-LHCC-2020-004, CMS-TDR-021",
    year        = "2020",
    url = {https://cds.cern.ch/record/2714892?ln=es},
}

@article{Zabi:2022,
    author  = "Zabi, Alexandre and {CMS Collaboration}",
    title   = "{System Design and Prototyping for the CMS Level-1 Trigger at the High-Luminosity LHC}",
    journal = "Journal of Physics: Conference Series",
    volume  = "2374",
    pages   = "012090",
    year    = "2022",
    doi     = "10.1088/1742-6596/2374/1/012090"
}

@misc{Folgueras:2025,
    author    = "Folgueras, Santiago",
    title     = "{INTREPID}",
    publisher = "European Dissemination",
    year      = "2025",
    doi       = {10.54050/PRJ2222858}
}

@inproceedings{Acevedo:2026,
    author    = "Acevedo, D. E. and Prado Pico, J. and Folgueras G{\'o}mez, S. and Vico Villalba, C. and Aller Guti{\'e}rrez, E. and {CMS Collaboration}",
    title     = "{Triggering on muon showers in the Barrel Muon Trigger of the CMS experiment for the HL-LHC upgrades}",
    booktitle = "Proceedings of Science (PoS)",
    volume    = "485",
    series    = "EPS-HEP2025",
    pages     = "577",
    year      = "2026",
    url       = "10.22323/1.485.0577"
}

@inproceedings{Leguina:2025,
    author = "Leguina, Pelayo and Folgueras, Santiago and Cardini, Andrea and Aller, Elena",
    title = "{Using AI on FPGAs for the CMS Overlap Muon Track Finder for the HL-LHC}",
    booktitle = "{2nd European AI for Fundamental Physics Conference}",
    eprint = "2509.23347",
    archivePrefix = "arXiv",
    primaryClass = "hep-ex",
    month = "9",
    year = "2025"
}

@article{Sirunyan:2020,
    author  = "Sirunyan, A. and others",
    title   = "{Performance of the reconstruction and identification of high-momentum muons in proton-proton collisions at $\sqrt{s} = 13$ TeV}",
    journal = "Journal of Instrumentation (JINST)",
    volume  = "15",
    pages   = "P02027",
    year    = "2020",
    doi={10.1088/1748-0221/15/02/P02027}
}

@article{Abbiendi:2023,
    author  = "Abbiendi, G. and others",
    title   = "{The Analytical Method algorithm for trigger primitives generation at the LHC Drift Tubes detector}",
    journal = "Nuclear Instruments and Methods in Physics Research Section A (NIMA)",
    volume  = "1049",
    pages   = "168103",
    year    = "2023",
    doi={https://doi.org/10.1016/j.nima.2023.168103}
}

@article{Prado:2026,
    author  = "Prado, J. and others",
    title   = "{Muon shower tagging in the barrel muon system of the CMS experiment}",
    journal = "Journal of Instrumentation (JINST)",
    volume  = "21",
    pages   = "C01050",
    year    = "2026",
    doi     = "10.1088/1748-0221/21/01/C01050"
}

@article{BUNKOWSKI:2019,
    title   = {The algorithm of the CMS Level-1 Overlap Muon Track Finder trigger},
    journal = {Nuclear Instruments and Methods in Physics Research Section A: Accelerators, Spectrometers, Detectors and Associated Equipment},
    volume  = {936},
    pages   = {368-369},
    year    = {2019},
    note    = {Frontier Detectors for Frontier Physics: 14th Pisa Meeting on Advanced Detectors},
    issn    = {0168-9002},
    doi     = {https://doi.org/10.1016/j.nima.2018.10.173},
    url     = {https://www.sciencedirect.com/science/article/pii/S0168900218314876},
    author  = {K. Bunkowski}
}

@article{Zabolotny:2016,
    author  = "Zabolotny, W. M. and Byszuk, A.",
    title   = "{Algorithm and implementation of muon trigger and data transmission system for barrel-endcap overlap region of the CMS detector}",
    journal = "Journal of Instrumentation (JINST)",
    volume  = "11",
    number  = "03",
    pages   = "C03004",
    year    = "2016",
    doi={10.1088/1748-0221/11/03/C03004}
}

@article{Hamilton:2017,
  author       = {William L. Hamilton and
                  Rex Ying and
                  Jure Leskovec},
  title        = {Inductive Representation Learning on Large Graphs},
  journal      = {CoRR},
  volume       = {abs/1706.02216},
  year         = {2017},
  url          = {http://arxiv.org/abs/1706.02216},
  eprinttype   = {arXiv},
  eprint       = {1706.02216},
  bibsource    = {dblp computer science bibliography, https://dblp.org}
}

@article{Iiyama:2020,
    AUTHOR={Iiyama, Yutaro  and Cerminara, Gianluca  and Gupta, Abhijay  and Kieseler, Jan  and Loncar, Vladimir  and Pierini, Maurizio  and Qasim, Shah Rukh  and Rieger, Marcel  and Summers, Sioni  and Van Onsem, Gerrit  and Wozniak, Kinga Anna  and Ngadiuba, Jennifer  and Di Guglielmo, Giuseppe  and Duarte, Javier  and Harris, Philip  and Rankin, Dylan  and Jindariani, Sergo  and Liu, Mia  and Pedro, Kevin  and Tran, Nhan  and Kreinar, Edward  and Wu, Zhenbin },
           
    TITLE={Distance-Weighted Graph Neural Networks on FPGAs for Real-Time Particle Reconstruction in High Energy Physics},
    JOURNAL={Frontiers in Big Data},
    VOLUME={Volume 3 - 2020},
    YEAR={2021},
    URL={https://www.frontiersin.org/journals/big-data/articles/10.3389/fdata.2020.598927},
  
DOI={10.3389/fdata.2020.598927},
  
ISSN={2624-909X},
}

@article{Sen:2008,
    author  = "Sen, P. and Namata, G. and Bilgic, M. and Getoor, L. and Gallagher, B. and Eliassi-Rad, T.",
    title   = "{Collective Classification in Network Data}",
    journal = "AI Magazine",
    volume  = "29",
    number  = "3",
    pages   = "93--106",
    year    = "2008",
    doi     = "10.1609/aimag.v29i3.2157"
}

@article{Jacob:2018,
  author       = {Benoit Jacob and
                  Skirmantas Kligys and
                  Bo Chen and
                  Menglong Zhu and
                  Matthew Tang and
                  Andrew G. Howard and
                  Hartwig Adam and
                  Dmitry Kalenichenko},
  title        = {Quantization and Training of Neural Networks for Efficient Integer-Arithmetic-Only
                  Inference},
  journal      = {CoRR},
  volume       = {abs/1712.05877},
  year         = {2017},
  url          = {http://arxiv.org/abs/1712.05877},
  eprinttype   = {arXiv},
  eprint       = {1712.05877},
  bibsource    = {dblp computer science bibliography, https://dblp.org}
}

@article{Syrunyan:2021,
  title         = {Search for long-lived particles using displaced jets in proton-proton collisions at $\sqrt{s}=13\text{ }\text{ }\mathrm{TeV}$},
  author        = {Sirunyan, A. M. and others},
  collaboration = {CMS},
  journal       = {Phys. Rev. D},
  volume        = {104},
  issue         = {1},
  pages         = {012015},
  numpages      = {38},
  year          = {2021},
  month         = {Jul},
  publisher     = {American Physical Society},
  doi           = {10.1103/PhysRevD.104.012015},
  url           = {https://link.aps.org/doi/10.1103/PhysRevD.104.012015}
}

@article{Cottin:2023,
   title     = {Long-lived heavy neutral leptons with a displaced shower signature at CMS},
   volume    = {2023},
   ISSN      = {1029-8479},
   url       = {http://dx.doi.org/10.1007/JHEP02(2023)011},
   DOI       = {10.1007/jhep02(2023)011},
   number    = {2},
   journal   = {Journal of High Energy Physics},
   publisher = {Springer Science and Business Media LLC},
   author    = {Cottin, Giovanna and Helo, Juan Carlos and Hirsch, Martin and Peña, Cristián and Wang, Christina and Xie, Si},
   year      = {2023},
   month     = Feb 
}

@article{Cardini:2026,
      title={Integer Quantization of Graph Neural Networks for Real-Time FPGA Track Finding}, 
      author={Andrea Cardini and Pelayo Leguina and Elena Aller and Santiago Folgueras},
      year={2026},
      eprint={2609.28144},
      archivePrefix={arXiv},
      primaryClass={hep-ex},
      url={https://arxiv.org/abs/2609.28144}, 
}

\end{document}